\documentclass{article}
\usepackage{amsmath}
\usepackage{amssymb}
\usepackage{times}
\newtheorem{Definition}{Definition}
\newtheorem{Lemma}{Lemma}

 \newcommand{\begeq}{\begin{equation}}
\newcommand{\bea}{\begin{eqnarray}}
\newcommand{\eea}{\end{eqnarray}} \newcommand{\nn}{\nonumber}

\newcommand{\ca}{$C^*$-algebra}

\newcommand{\sth}{$\mbox{}^*$-homomorphism}

 \newcommand{\til}{\tilde}
\newcommand{\raw}{\rightarrow}

\newcommand{\rst}{\upharpoonright} 
\newcommand{\x}{\times} 
 
\newcommand{\cin}{C^{\infty}} \newcommand{\cci}{C^{\infty}_c}

\newcommand{\inv}{^{-1}}

\newcommand{\al}{\alpha} \newcommand{\bt}{\beta}
\newcommand{\gm}{\gamma} \newcommand{\Gm}{\Gamma}
 
\newcommand{\ep}{\epsilon}

 \newcommand{\sg}{\sigma}

\newcommand{\CO}{{\mathcal O}}

 \newcommand{\R}{{\mathbb R}}
 
 \newcommand{\SG}{\mathsf{G}}

  \makeatletter
\newskip\tempskip \def\endproof{{\parfillskip24\p@ plus\@ne
fil\@@par}\tempskip\prevdepth
\ifdim\lastskip=\z@\tempskip\z@\else\vskip-\lastskip
\ifdim\tempskip>4\p@ \tempskip.5\tempskip \else \tempskip\z@\fi\fi
\nobreak\vskip-\baselineskip\vskip-\tempskip\noindent\hbox
to\hsize{\hfill
$\blacksquare$}\par\vskip\tempskip\vskip\abovedisplayskip\@doendpe}
\makeatother \makeatletter
\newskip\tempskip \def\endiproof{{\parfillskip24\p@ plus\@ne
fil\@@par}\tempskip\prevdepth
\ifdim\lastskip=\z@\tempskip\z@\else\vskip-\lastskip
\ifdim\tempskip>4\p@ \tempskip.5\tempskip \else \tempskip\z@\fi\fi
\nobreak\vskip-\baselineskip\vskip-\tempskip\noindent\hbox
to\hsize{\hfill
$\Box$}\par\vskip\tempskip\vskip\abovedisplayskip\@doendpe}
\makeatother \newcommand{\enp}{\endproof}

\newcommand{\BCC}{Baum--Connes conjecture}
\begin{document}
\pagestyle{plain}
\title{Deformation quantization and the Baum--Connes conjecture\thanks{Dedicated to Rudolf
Haag at his 80th birthday. }}
\author{\textsc{N.P. Landsman}\thanks{Supported by  Stichting FOM.}\\
Korteweg--de Vries Institute for Mathematics\\
University of Amsterdam\\
Plantage Muidergracht 24\\
NL-1018 TV AMSTERDAM \\ THE NETHERLANDS\\
email \texttt{npl@science.uva.nl}}
\date{today}
\maketitle
\begin{abstract}
 Alternative titles of this paper would have been `Index theory
 without index' or `The Baum--Connes conjecture without Baum.'

  In
 1989, Rieffel introduced an analytic version of deformation
 quantization based on the use of continuous fields of
 $C^*$-algebras. We review how a wide variety of examples of such quantizations can
 be understood on the basis of a single lemma involving amenable
 groupoids. These include Weyl--Moyal quantization on manifolds,
 $C^*$-algebras of Lie groups and Lie groupoids, and the E-theoretic
 version of the Baum--Connes conjecture for smooth groupoids as
 described by Connes in his book \textit{Noncommutative Geometry}.

 Concerning the latter, we use a different semidirect product
 construction from Connes. This enables one to formulate the
 Baum--Connes conjecture in terms of twisted Weyl--Moyal
 quantization. The underlying mechanical system is a noncommutative
 desingularization of a stratified Poisson space, and the \BCC\
 actually suggests a strategy for quantizing such singular spaces.
\end{abstract}
\section{Introduction}
As a tribute to Rudolf Haag, this paper is a double
provocation. Firstly, it is about quantization, a concept Haag
apparently doesn't like. Indeed, he has always stressed that (local)
quantum physics stands on its own, and should not be thought of as the
quantization of some classical theory. Secondly, it fits in the
ideology of `physical mathematics,' in attempting to understand a
concept in pure mathematics (viz.\ the \BCC), in terms of ideas from physics (namely
quantization).  Characteristically, there is not a single theorem in
this paper.  As the founding editor of \textit{Communications in
Mathematical Physics}, Haag may well have second thoughts about the
seemingly irrepressible development of his journal into a medium for
both `mathematical physics' and `physical mathematics.'  On the
positive side, concerning the first point we use a formulation of
quantization in terms of \ca s, and even manage to relate the \BCC\ to the algebraic
theory of superselection rules initiated by Haag \cite{Haa}.
With regard to the
second, we note that this paper only contains valid mathematics, which
everyone can check and understand.

One source of inspiration for this paper is the known relationship
between index theory (in the sense of Atiyah and Singer \cite{AS1})
and quantum physics. This relationship was discovered in the context
of anomalies in quantum field theory \cite{ASZ,AGP}, and is closely
related to supersymmetry \cite{AG1,FW}. See, e.g., \cite{BGV,Bis,Vor}
for representative mathematical
literature generated by this line of research. 
On a different note, it turns out that index theory is
closely linked to deformation quantization \cite{ENN2,Fed,NT1}.
It remains unclear (at least to the author) how the supersymmetric
approach to index theory is related to the one based on deformation theory.

A promising way of  looking  at the relationship between quantization and
index theory is to involve the K-theory of \ca s
\cite{Bla,Ror}. Pragmatically speaking, K-theory  is
the (generalized) cohomology theory  of
algebraic topology that is best adapted to a generalization  to
noncommutative \ca s. K-theory is defined by functors
$K_n$, $n\in\mathbb{Z}$, from \ca s to abelian groups, which are
stable, homotopy invariant, and satisfy Bott periodicity
$K_{n+2}(A)\cong K_n(A)$ (natural in $A$). One therefore
simply writes the K-theory of $A$ as $K_*(A)$, where $*=1,2$.
 Bott periodicity  leads to a periodic
(or cyclic) 6-term exact sequence associated to a short exact sequence, 
which underlies most explicit  computations in K-theory.  

K-theory for \ca s is a fundamental tool in noncommutative geometry
\cite{Con}, and also plays a key role in Elliott's classification program
for simple nuclear \ca s \cite{Lin,RS}. In mathematical physics, the
best-known applications of noncommutative K-theory have been to the
theory of the quantum Hall effect \cite{Bel} and to the description
of quasi-crystals \cite{KP}. So far, the use of K-theory in physical
mathematics seems limited to the commutative case \cite{WitK}.

The bivariant E-theory of Connes and Higson \cite{Con,CH}
is a generalization of the K-theory of \ca s, which at the same time
provides maps between K-groups. E-theory is based on specific deformations
of \ca s, and is closely related to index theory \cite{Con,Hig}.
Thus it seems natural to use E-theory in an attempt to further clarify the
relationship between index theory and quantization. However,
deformation quantization contains an ingredient that seems to be missing
in E-theory, namely the Poisson bracket. This determines the `direction'
of a deformation, providing information that could be useful in understanding
why certain maps between K-groups defined by E-theory occur naturally. 
Indeed, this is a guiding thought behind this paper.

One of the main issues in K-theory in the context of noncommutative geometry
is the so-called \BCC, which is closely related to index theory
\cite{BC,BCH,Con} (in this paper, we restrict ourselves to the
conjecture ``without coefficients''). Here the problem is to give a
geometric description of the K-theory of the reduced \ca\ $C_r^*(G)$
of a group \cite{Dix,Ped} or groupoid \cite{Ren} $G$.  This is potentially interesting for
physics, since algebras of observables of a large class of quantum
mechanical systems are of the form  $C_r^*(G)$ \cite{Lan}, and the K-theory of
such algebras is an invariant of the physical description that
deserves to be explored.

For a compact group, $K_0(C_r^*(G))$ equals the  free abelian group on $\hat{G}$
(the unitary dual of $G$, which in this case is discrete), whereas $K_1$ is trivial.
The groupoid analogue of a compact group is a proper groupoid;
a groupoid $G$
with base $G^{(0)}$ and source and range maps $s,r: G\raw G^{(0)}$, respectively,
is called proper when $(r,s):G\raw G^{(0)}\x G^{(0)}$ is a proper map.
The K-theory  of the reduced \ca\ of such a groupoid
can in principle be described in terms 
of the K-theory of the compact stability groups $G_u^u$ of points
$u\in G^{(0)}$ \cite{Ren},  combined with the (equivariant) topological K-theory of the
orbit space $G^{(0)}/G$ (which is locally compact and Hausdorff).
Hence the compact or proper case is fully understood in principle. 

One idea behind the \BCC\ is to `tame' a noncompact group or nonproper
groupoid by letting it act properly on some space. Under a proper
action, all stability groups are compact, and the orbit space is
locally compact and Hausdorff \cite{AR}.  Baum and Connes define a computable
topological K-theory $K^*_{\mathrm{top}}(G)$ in terms of such proper
actions, and relate it to the actual K-theory $K_*(C_r^*(G))$ by a map
$\mu$, called the analytic assembly map. The \BCC\ states that $\mu$ should be 
an isomorphism. This would, then, render $K_*(C_r^*(G))$ computable as well.

The \BCC\ actually enjoys a number of different formulations. For
groups, the standard version is that of \cite{BCH}. Here
$K^*_{\mathrm{top}}(G)$ is defined in terms of the $G$-equivariant
K-homology of the classifying space $\underline{E}G$ of $G$ for proper
actions. Roughly speaking, elements of $K^*_{\mathrm{top}}(G)$ are
equivalence classes of $G$-invariant operators on some Hilbert space
carrying representations of $G$ as well as of $C_0(X)$, where $X$ is
some proper $G$ space. These operators have an index taking values in
$K_*(C^*(G))$, and $\mu$ is essentially this index, composed with the
K-theory map $\pi_{r*}$ induced by the canonical projection
$\pi_r:C^*(G)\raw C^*_r(G)$.  Thus the \BCC\ states that, in a
suitably injective way, every element of $K_*(C_r^*(G))$ may be
represented as an index.

In this form, the \BCC\ has been proved for large classes of discrete 
or algebraic groups
(cf.\ \cite{Ska,Val}), as well as for all almost connected locally compact
groups \cite{CEN}.  There exists an analogous formulation for
locally compact groupoids with Haar system, surveyed in \cite{Tu}.
The usual formulation of the \BCC\ for both groups and groupoids 
is based on Kasparov's KK-theory (cf.\ \cite{Bla}), which is also the
fundamental tool in the extant proofs of special cases of the conjecture.

A different approach to the \BCC, based on E-theory,
 was initiated by Connes himself
\cite[\S II.10]{Con}. The main purpose of
this paper is to make explicit how Connes's E-theoretic formulation 
of the \BCC\ is nothing but the
statement that the $G$-twisted  Weyl--Moyal quantization of a
certain space preserves K-theory. This is actually closely
related to Connes's own way of seeing the \BCC\ as a $G$-equivariant
version of Bott periodicity.  To accomplish this, we have to slightly
modify Connes's construction of the analytic assembly map $\mu$ in order to bring it
in line with the \ca ic approach to Weyl--Moyal
quantization. Moreover, we prove a fundamental and nontrivial
continuity property left to the reader in \cite{Con}.
As suggested above, the use of deformation quantization amplifies
E-theory by providing the direction of the deformation defining $\mu$.

When $G$ is a Lie group,  the classical mechanical systems underlying the above approach to the \BCC\ are Poisson spaces of the type $T^*(P)/G$, where $P$ is a proper $G$ space,
and the $G$ action on $T^*(P)$ is  the pullback of the one on $P$. This action
automatically preserves the canonical Poisson bracket (or, equivalently,
the symplectic form) on $T^*(P)$, which therefore descends  to a
 Poisson structure on  $T^*(P)/G$. 
 In case that $P$ is a principal $G$ bundle (i.e., when the $G$ action is free),
$T^*(P)/G$ is a manifold, whose physical interpretation is well understood
in terms of a particle moving on the configuration space $Q=P/G$, coupled to
an external gauge field \cite{Mar}. The algebra of observables of the
corresponding quantum system \cite{Lan} is the \ca\ of the so-called gauge groupoid
$(P\x P)/G$ of the  principal $G$ bundle $P$ \cite{Mac}. Such a quantum system has
a nontrivial superselection structure, which is fully described by the irreducible
unitary representations of $G$. Similarly, the underlying classical system 
has `classical superselection sectors,' defined as
the  symplectic leaves of
$T^*(P)/G$ \cite{Mar}. In analogy to the quantum situation, these turn out 
to correspond to the coadjoint orbits of $G$.

However, when the $G$ action on $P$ is not free (and this is the main case of
interest in connection with the \BCC), the quotient $T^*(P)/G$ is no longer
a manifold. In fact, the \BCC\ for Lie groups \`{a} la Connes is formulated 
in terms of a noncommutative desingularization of $T^*(P)/G$, namely the
crossed product \ca\ $C_0(T^*(P))\rtimes G$. The structure of $T^*(P)/G$
as a singular space is well known \cite{LMS}: its naive symplectic leaves
are actually stratified symplectic spaces \cite{SL}, which 
further decompose as unions of symplectic manifolds. This introduces
additional classical superselection sectors, which should be related
to the structure of the desingularization $C_0(T^*(P))\rtimes G$ in some
way. In any case, inspired by Connes's E-theoretic formulation of the \BCC, we
are led to a concrete proposal to quantize the singular Poisson space $T^*(P)/G$
by deforming its noncommutative desingularization. 
 
The plan of this paper is as follows. In Section \ref{bs} we review
the notion of \ca ic deformation quantization, and state the key
technical lemma, on which most subsequent arguments will be based. In
Section \ref{examples} we discuss a number of examples relevant to the
\BCC, and in Section \ref{BCC} we turn to the \BCC\ itself.
Finally, in Section \ref{pi} we provide the details of the physical
interpretation sketched above.

We hope that this expository paper attracts mathematical physicists 
to the \BCC, and draws the attention of noncommutative geometers to
the problem of quantizing singular symplectic spaces \cite{Proc}.
\medskip

\textbf{Acknowledgements} The author is greatly indebted to Rudolf Haag
for his profound intellectual influence, and wishes to take this
opportunity to also thank Haag's pupils and collaborators Detlev
Buchholz, Klaus Fredenhagen, and Daniel Kastler for their generous
hospitality and support in various phases of his scientific career.

This paper is based on lectures invited by  B. Monthubert
in the S\'{e}minaire M\'{e}diterran\'{e}en
d'Alg\`{e}bre et Topologie in Toulouse, March 2002, and at the
Oberwolfach meeting on Noncommutative Geometry, April 2002
at the invitation of A. Connes, J. Cuntz, and M. Rieffel. 
It is a pleasure to thank many participants of these meetings
for comments and criticism. 
\section{Basic setting}\label{bs}
The $C^*$-algebraic approach to deformation quantization was initiated
in 1989 by Rieffel \cite{Rie1}, who observed that a number of examples
of quantization could be described by continuous fields of
$C^*$-algebras in a natural and attractive way. We refer to
\cite{Lan,Rie2} for surveys of the starting period of $C^*$-algebraic
deformation quantization, including references. 

We now review the basic definitions pertinent to \ca ic deformation quantization.
On the classical side, we have
\begin{Definition}
A Poisson algebra is a commutative algebra $\til{A}$ over $\mathbb{C}$
equipped with a Lie bracket $\{\, ,\,\}$, such that
 for each $f\in \til{A}$ the map $g\mapsto \{f,g\}$
is a derivation of $\til{A}$ as a commutative algebra.
A Poisson manifold $P$ is a manifold equipped with a Lie bracket on
 $\til{A}=\cin(P)$, such that it becomes  a Poisson algebra
with respect to  pointwise multiplication.
\end{Definition}

On the quantum side, one needs
\begin{Definition}\label{Defcf}
A field of \ca s over a compact Hausdorff space $X$ is a triple $(X,
\{A_x\}_{x\in X}, A)$, where $\{A_x\}_{x\in X}$ is some family of \ca
s indexed by $X$, and $A$ is a family of sections (that is, maps
$a:X\raw \coprod_{x\in X}A_x$ for which $a(x)\in A_x$) that is:
\begin{enumerate}
\item
 A \ca\ under pointwise operations and the natural norm $$\|
a\|=\sup_{x\in X} \| a(x)\|_{A_x};$$ 
\item Closed under
multiplication by $C(X)$;
\item Full, in that for each $x\in X$ one has $\{a(x)\mid a\in
A\}=A_x$.
\end{enumerate}  The field is said to be continuous when for each $a\in A$
the function $x\mapsto \| a(x)\|$ is in $C(X)$.
\end{Definition}

This is equivalent to the corresponding definition of Dixmier
\cite{Dix}; cf.\ \cite{Blan,KW}. Such a field comes with a collection
of \sth s $\pi_x:A\raw A_x$, defined by $\pi_x(a)=a(x)$. 
We will use this  with $X=I=[0,1]$, 
seen as the set of values of Planck's constant $\hbar$.

Poisson manifolds are related to continuous fields through the concept of
\ca ic deformation quantization. 
\begin{Definition}\label{gsq}
A \ca ic deformation quantization of a Poisson manifold $P$ consists of:
\begin{enumerate}
\item A
continuous field of \ca s $(I, \{A_{\hbar}\}_{\hbar\in I},A)$ in which
$A_0=C_0(P)$;
\item  A Poisson subalgebra $\til{A}_0$ of $\cin(P)$ that is densely
contained in $C_0(P)$;
\item  A cross-section $Q:\til{A}_0\raw A$ of $\pi_0$, 
\end{enumerate}
such that, in terms of $Q_{\hbar}=\pi_{\hbar}\circ Q$,  for all $f,g\in \til{A}_0$ one has
\begin{equation}
\lim_{\hbar\rightarrow 0} 
\|\frac{i}{\hbar}[Q_{\hbar}(f),Q_{\hbar}(g)]-Q_{\hbar}(\{f,g\})\|_{\hbar} =0. \label{Dirac}
\end{equation}
\end{Definition}

The idea behind (\ref{Dirac}), which may be traced  back to Dirac, is
that the Poisson bracket on $P$ determines the direction in which
$C_0(P)$ is deformed into a noncommutative \ca. 
In any case, this definition (with evident modifications when $I=[0,1]$ is replaced by
a more general index set) seems to cover practically all known examples.

A surprisingly large collection of examples can be constructed from the following data
\cite{LR,Ram}. We refer to \cite{Mac,Ren} for the theory of groupoids.
 Recall that a Lie groupoid is a groupoid where all
spaces and maps are smooth, and $s$ are $r$ are surjective submersions
\cite{Mac}.
\begin{Definition}\label{field}
A field of groupoids is a triple $(\SG,X,p)$, with $\SG$ a groupoid,
 $X$ a set, and $p:\SG\rightarrow X$ a surjection  such that $p=p_0\circ r=p_0
\circ s$, where $p_0=p\rst\SG^{(0)}$. If $\SG$ is a locally compact groupoid
and $X$ is a topological space, one requires that $p$ is continuous
and open.  When $\SG$ is a Lie groupoid and $X$ a manifold, $p$ should
be a surjective submersion. 
\end{Definition}

It follows that each $\SG_x=p\inv(x)$ is a subgroupoid of $\SG$
over  $\SG^{(0)}\cap p\inv(x)$, so that
$\SG=\coprod_{x\in X}
\SG_x$ as a groupoid. This holds algebraically, topologically, or
smoothly, as appropriate.

In the context of deformation quantization,
the following two cases occur: either $\SG$ is smooth, or $\SG$ is
\'{e}tale.  In both cases, $\SG$ and all $\SG_x$ automatically have a
(left or right) Haar system
\cite{Lan,LR,Ram,Ren}.  More generally, one may simply assume
that $\SG$ is a locally compact groupoid with Haar system.  One may then
form the convolution \ca s $C^*(\SG)$ and $C^*(\SG_x)$, or the
corresponding reduced \ca s $C_r^*(\SG)$ and $C_r^*(\SG_x)$
\cite{Con,Ren}. Each $a\in
C_c(\SG)$ (or $\cci(\SG)$) defines $a_x=a\rst \SG_x$ as an element of
$C_c(\SG_x)$ (etc.). These maps $C_c(\SG)\raw C_c(\SG_x)$ are continuous in
the appropriate norms, and extend to maps $\pi_x: C^*(\SG)\raw
C^*(\SG_x)$.  Hence one obtains a field of \ca s
$$(X,\{A_x=C^*(\SG_x)\}_{x\in X},A=C^*(\SG)),$$ where $a\in C^*(\SG)$
defines the section $x\mapsto \pi_x(a)$. A similar statement applies
to the corresponding reduced \ca s.

 The question now arises when this field is continuous. The answer,
generalizing certain results by Rieffel for groups \cite{Rie0}, is as follows.
\begin{Lemma}\label{ramlem}
The field  $(X,\{C^*(\SG_x)\}_{x\in X},C^*(\SG))$ is continuous at all points
where $\SG_x$ is amenable \cite{AR,Ren} (and similarly for the case of
reduced \ca s).
\end{Lemma}

This lemma was first mentioned to the author by Skandalis in 1997; see
\cite[p.\ 469]{Lan}.  A complete proof, based on results of Skandalis's student 
Blanchard \cite{Blan}, appeared in \cite{Ram}, and was repeated in
\cite{LR}.  In our examples of deformation quantization, where $X=I$,
two possibilities occur. 

In the first situation,  all $\SG_{\hbar}$ are amenable, in
which case Lemma \ref{ramlem} immediately proves continuity of the
field in question. See \cite{Cad} for a description of the noncommutative
tori of Rieffel \cite{Rie1} and of the noncommutative four-spheres of Connes and
Landi \cite{CL} (and of many other examples) as deformation quantizations
along these lines.

In the second situation, typically only $\SG_0$ is amenable, and the
field is trivial away from $\hbar=0$ (see below). The former property
then yields continuity at $\hbar=0$ by the lemma, whereas the latter
gives continuity on $(0,1]$. In the context of Definition \ref{gsq},
the reason why $G_0$ is amenable is that  $A_0$ must be commutative,
which implies that $G_0$ is a bundle of abelian groups. But such
groupoids are always amenable \cite{AR}.  In both cases, one obtains a
continuous field.

Here a continuous field $(I,\{A_{\hbar}\}_{\hbar\in I},A)$ is said to be trivial away from
$\hbar=0$ when $A_{\hbar}=B$ for all $\hbar\in (0,1]$,
and one has a  short exact sequence 
\begin{equation}
0\raw CB\raw A\raw A_0\raw 0, \label{SES}
\end{equation}
in terms of the so-called cone $CB=C_0((0,1],B)$. For later use, we
recall that such a field induces a map $K_*(A_0)\raw K_*(B)$
in the following way \cite{Con,CH}. 
Since the cone $CB$  is contractible, and therefore has trivial
K-theory,  the periodic 6-term sequence  shows that
\begin{equation}
\pi_{0*}: K_*(A)\raw K_*(A_0) \label{Kiso}
\end{equation}
is an isomorphism; here $\pi_{0*}$ stands for the image of 
the $\mbox{}^*$-homomorphism $\pi_0:A\raw A_0$ under the K-functor.
The K-theory map defined by the  field is then simply
\begin{equation}
\pi_{1*}\circ\pi_{0*}\inv: K_*(A_0)\raw K_*(B). \label{Kmap}
\end{equation}
\section{Examples}\label{examples}
\subsection{Particle on a manifold}\label{3.1}
The simplest physically relevant example of this setting is provided
by Connes's tangent groupoid $G_M$ of a manifold $M$; see \cite[p.\
102]{Con}.  Here $$\SG=G_M=\coprod_{\hbar\in I} G_{\hbar},$$ where
$\SG_0=T(M)$ is the tangent bundle of $M$, seen as a groupoid over $M$
under addition in each fiber, and $\SG_{\hbar}=M\x M$ for all
$\hbar\in (0,1]$ is the pair (or coarse) groupoid on $M$. The point
is, of course, that $\SG$ has a smooth structure turning it into a Lie
groupoid (see below).

The corresponding field of \ca s has fibers
\begin{eqnarray}
A_0 & = & C^*(T(M))\cong C_0(T^*(M)) ;\nn \\
A_{\hbar} & = & C^*(M\x M)\cong B_0(L^2(M))\:\: \forall\hbar\in(0,1], \label{CF1}
\end{eqnarray}
where $B_0(H)$ is the \ca\ of compact operators on $H$.  
For later use, it is crucial to remark that the isomorphism in the first equation
is given by a fiberwise Fourier transformation. The
continuity of this field follows from Lemma \ref{ramlem} as explained
above (among many other proofs; cf.\ \cite{ENN1,Lan} and references
therein).  For the quantization maps $Q_{\hbar}$ see \cite{Lan,LGCA,Pflaum};
these are essentially given by Weyl--Moyal quantization with respect
to a Riemannian structure on $M$. 
The relationship between the tangent groupoid and quantization was
independently noted by Connes during his lectures at Les Houches in 1995;
see \cite{CCFGRV}.

Combining the trace $\mathrm{tr}$ (to implement the
isomorphism $K_0(B_0)\cong\mathbb{Z}$) with the map in (\ref{Kmap}),
one obtains a map
\begin{equation}
\mathrm{ind}_a=\mathrm{tr}\circ\pi_{1*}\circ\pi_{0*}\inv: K^0(T^*(M))\raw \mathbb{Z},  \label{inda}
\end{equation}
which is precisely 
the analytic index of Atiyah and Singer \cite{AS1}; cf.\  Lemma II.5.6 in \cite{Con}.
For $M=\mathbb{R}^n$, one has $K^0(\mathbb{R}^{2n})\cong \mathbb{Z}$, and
the analytic index   is the isomorphism $\bt$ of the Bott periodicity theorem 
\cite{At1}. 
The fact that the ``classical algebra'' $C_0(\mathbb{R}^{2n})$ and
the ``quantum algebra'' $B_0(L^2(\R^n))$ have the same
K-theory is peculiar to this special case; for general $M$
this will, of course, fail. The special case $M=\R^n$, however, lies behind
the Baum--Connes conjecture; see below. 
\subsection{Particle with internal degree of freedom}\label{3.2}
The above example describes the quantization of a particle moving on $M$,
with phase space $T^*(M)$. If, on the other hand, a particle has no kinematic
degrees of freedom (in that it does not move on a configuration space),
 but is only endowed with internal degrees of freedom, described
by a Lie group $G$, its algebra of observables is the group \ca\ $C^*(G)$.
As first recognized in \cite{Rie4} (under certain assumptions, which later turned out
to be unnecessary \cite{LR,Ram}), this algebra is a deformation quantization 
in the sense of Definition \ref{gsq} of
the Poisson manifold $\mathfrak{g}^*$, where $\mathfrak{g}$ is the Lie algebra
of $G$, and its dual vector space $\mathfrak{g}^*$ is equipped with the so-called
Lie--Poisson structure (which on linear functions is just given by the Lie bracket)
\cite{Lan,Mar}. 

The underlying Lie groupoid $\SG$ has fibers $\SG_0=\mathfrak{g}$ and
$\SG_{\hbar}=G$ for $\hbar\in (0,1]$. Here $\mathfrak{g}$ is regarded as
an abelian group, so that it is amenable, and Lemma \ref{ramlem}
proves continuity of the associated field of \ca s
\begin{eqnarray}
A_0 & = & C^*(\mathfrak{g})\cong C_0(\mathfrak{g}^*) ;\nn \\
A_{\hbar} & = &  C^*(G)  \:\: \forall\hbar\in(0,1]. \label{CF2}
\end{eqnarray}
Here $\mathfrak{g}$ is treated as an abelian group;
 once again,  the isomorphism in the first equation
is given by the Fourier transformation. 
The quantization maps $Q_{\hbar}$ are defined in terms of the usual exponential
map from $\mathfrak{g}$ to $G$, and  Definition \ref{gsq} turns out to be
satisfied.
\subsection{The Connes--Mackey semidirect product deformation}
The deformation described by Connes in \cite[p.\ 141]{Con} is similar
to the preceding example, with the difference that only the `noncompact
part' of $G$ is deformed.
Let $G$ be a connected Lie group with maximal compact subgroup $H$.
With  $\mathfrak{m}=T_e(G/H)$ one has $\mathfrak{g}=\mathfrak{h}\oplus
\mathfrak{m}$, and $H$ acts naturally on $\mathfrak{m}$.
One then has a Lie groupoid $\SG$ that is a field of groups with fibers
$\SG_0=\mathfrak{m}\rtimes H$ and $\SG_{\hbar}=G$ for $\hbar\in(0,1]$.
Since $\mathfrak{m}\rtimes H$ is amenable, Lemma \ref{ramlem} proves
continuity of the associated field of \ca s. Note that, unlike in
the previous examples, $A_0=C^*(\mathfrak{m}\rtimes H)$ is now noncommutative, 
  like $A_{\hbar}=C^*(G)$ (except in trivial cases). 
\subsection{Poisson manifolds associated to Lie algebroids}\label{3.4}
Examples \ref{3.1} and \ref{3.2} are both special cases of a very general
construction \cite{Lan,LGCA,LR,Ram}. A Lie algebroid $E$ is a (real) vector
bundle over a manifold $M$, whose space $\Gm(E)$ of smooth sections is
equipped with a Lie bracket satisfying the Leibniz rule
\begin{equation}
 [s_1, fs_2]=f[s_1,s_2]+ (\al\circ s_1)f\cdot s_2 \label{LR}
\end{equation}
 for some vector bundle map $\al:E\raw T(M)$. Such a map,
called the anchor map of the Lie algebroid, is unique when it exists.
(This definition, which we learnt from Marius Crainic, is more efficient 
than the usual one \cite{CW,Lan,Mac}.) The simplest example is
$E=T(M)$, where $\al$ is the identity map. 

A Lie groupoid $G$ has an associated Lie algebroid $A(G)$ over the
base space $G^{(0)}$ \cite{CW,Lan,Mac}. The dual vector bundle
$A^*(G)$ has a canonical Poisson structure, which generalizes both the
usual symplectic structure on $T^*(M)$ and the Lie--Poisson bracket on
$\mathfrak{g}^*$ \cite{CDW,Cou}. Generalizing Connes's tangent
groupoid \cite{HS,Wei89} (which emerges as a special case for $G=M\x
M$), there exists a Lie groupoid $\SG=\coprod_{\hbar\in I} G_{\hbar}$,
where $\SG_0=A(G)$ (seen as a Lie groupoid over $G^{(0)}$ under
addition in each fiber) and $\SG_{\hbar}=G$ for $\hbar>0$.
With abuse of terminology, this is called the tangent groupoid of $G$.

As noted in \cite{NWX}, the Lie algebroid of $\SG$ is the so-called
adiabatic Lie algebroid associated to $A(G)$. In general, the
adiabatic Lie algebroid $E_t$ associated to some Lie algebroid $E$
over $M$ is a vector bundle over $M\x I$ whose total space is the
pullback $\mathrm{pr}_1^*E$ of the map $\mathrm{pr}_1:M\x I\raw M$; the Lie bracket
is, in obvious notation,
\begin{equation} 
[s_1,s_2]_{E_t}(\cdot,\hbar)=\hbar [s_1(\hbar),s_2(\hbar)]_E.
\end{equation}
The tangent groupoid of $G$ is then obtained by applying the integration
procedure of \cite{CF} to $A(G)_t$; this provides, in particular, the smooth
structure.

By our standard argument, the associated field of \ca s
\begin{eqnarray}
A_0 & = & C^*(A(G))\cong C_0(A^*(G)) ;\nn \\
A_{\hbar} & = &  C^*(G)  \:\: \forall\hbar\in(0,1], \label{CF3}
\end{eqnarray}
 is continuous, and provides a deformation quantization of the Poisson
 manifold $A^*(G)$ in the sense of Definition \ref{gsq}.  As in the previous examples,
 the isomorphism in the first equation is given by a fiberwise Fourier
 transformation. The analogy between the maps $G\mapsto A^*(G)$ and
$G\mapsto C^*(G)$ is quite deep; see \cite{NPLOA}.
\subsection{Gysin maps}
Certain constructions of Connes in index theory turn out to be special cases of
Example \ref{3.4}. One instance is the  `shriek' map 
$p!: K^*(F^*)\raw K_*(C^*(V,F))$ on
p.\ 127 of \cite{Con}, which plays a key role both in the longitudinal
index theorem for foliations and in the construction  of the
analytic assembly map for foliated manifolds. Here $V$ is a manifold with foliation
$F\subset T(V)$, and $C^*(V,F)=C^*(G(V,F))$ is  the canonical \ca\ of the holonomy groupoid $G(V,F)$ of the foliation.
Now $p!$  is nothing but the K-theory map
(\ref{Kmap}) induced by the continuous field (\ref{CF3}),  where $G=G(V,F)$.  
 The analytic index (\ref{inda}) corresponds to the special case that $V=M$ is trivially
foliated (i.e., $F=T(M)$). 

 The index groupoid defined in \cite[\S II.6]{Con} is another example of (\ref{Kmap}) with
 (\ref{CF3}).  Let  $L:E\raw F$ be a vector bundle map between vector
bundles over a common base $B$. Then  one has a Lie groupoid $G=\mathrm{Ind}_L=F\rtimes_L
 E$ over $F$, whose Lie algebroid is $F\x_B E$. The latter is a vector
 bundle over $B$, and in the formalism of this paper it should be regarded as
 a groupoid over $F$ under addition in each fiber. Hence
 $A_0=C^*(F\x_B E)\cong C_0(F\x E^*)$. The corresponding map
 (\ref{Kmap}) is basic to  Connes's construction of the Gysin or shriek map $f!:
 K^*(X)\raw K^*(Y)$ induced by a smooth K-oriented map $f:X\raw Y$ between
 two manifolds.
\section{The \BCC}\label{BCC}
We first recall a generalized semidirect product construction for
groupoids, which is necessary to relate the \BCC\ to quantization.
We then describe the analytic assembly map \`{a} la Connes. 
In what follows, $G$ is a Lie groupoid over $G^{(0)}$.
\subsection{On semidirect products}
 Recall \cite{Con,Mac} that a (right) $G$ space $P$ is a
smooth map $P\stackrel{\al}{\raw}G^{(0)}$ along with a map $P\x_{G^{(0)}}
G\raw P$, where
\begin{equation}
P\x_{G^{(0)}} G=\{(p,\gm)\in P\x G\mid
\al(p)=r(\gm)\},\label{PG}
\end{equation} written as
$(p,\gm)\mapsto p\gm$, such that
$(p\gm_1)\gm_2=p(\gm_1\gm_2)$ whenever defined, $p\al(p)=p$ for all
$p$, and $\al(p\gm)=s(\gm)$. The action is called proper when $\al$ is
a surjective submersion and the map $P\x_{G^{(0)}} G\raw P\x P$,
$(p,\gm)\mapsto (p,p\gm)$ is proper (in that the inverse images of
compact sets are compact). 

In Connes's description of the \BCC\ \cite{Con}, the standard
semidirect product construction in groupoid theory is used: if $G$
acts on a space $P$ as above, one forms a groupoid $P\rtimes G$ over
$P$, with total space $P\x_{G^{(0)}} G$, source and range maps $s(p,\gm)=p\gm$ and
$r(p,\gm)=p$, inverse $(p,\gm)\inv=(p\gm,\gm\inv)$, and multiplication
$(p,\gm)\cdot (p\gm,\gm')= (p,\gm\gm')$. However, as we shall see
shortly, the use of these semidirect products distorts the
relationship between the \BCC\ and deformation quantization.  For 
our purposes, we must 
work with generalized semidirect products (see \cite{AR} for the locally compact case and
\cite{Mac} (2nd ed.) for the smooth case).
 
 Let a $G$ space $H$
be a Lie groupoid itself, and suppose the base map $H\stackrel{\al}{\raw}G^{(0)}$ 
is a surjective submersion that satisfies 
\begin{enumerate}
\item $\al_0\circ s_H=\al_0\circ r_H=\al$
(cf.\ Definition \ref{field}); in other words, $H$ is a field of
groupoids over $G^{(0)}$, and $\al$ is a morphism of groupoids if
$G^{(0)}$ is seen as a space (where a groupoid $X$ is a space when
$X^{(0)}=X$ and $s=r=\mathrm{id}$).
\item For each $\gm\in G$, the map $\al\inv(r(\gm))\raw \al\inv(s(\gm))$,
$h\mapsto h\gm$, is an isomorphism of Lie groupoids; note that for each
$u\in G^{(0)}$, $\al\inv(u)$ is a Lie groupoid over $\al\inv(u)\cap H^{(0)}$.
In other words, one has $(h_1h_2)\gm=(h_1\gm)(h_2\gm)$ whenever defined.
\end{enumerate}
 
Under these conditions, one may define a Lie groupoid $H\rtimes G$,
 called the generalized semidirect product
of $H$ and $G$. The total space of $H\rtimes G$ is $H\x_{G^{(0)}} G$
as in (\ref{PG}),
the base space  $(H\rtimes G)^{(0)}$is $H^{(0)}$, the source and range
maps are 
\begin{eqnarray}
s(h,\gm) & = & s_H(h)\gm; \nn \\
r(h,\gm) & = & r_H(h), \label{sr}
\end{eqnarray}
respectively, the inverse is $(h,\gm)\inv =(h\inv\gm,\gm\inv)$ (note
that one automatically has $\al(h\inv)=\al(h)$, so that this element
is well defined), and multiplication is given by
$(h_1,\gm_1)(h_2\gm_1,\gm_2)=(h_1h_2,\gm_1\gm_2)$, defined whenever
the product on the right-hand side exists (this follows from the
automatic $G$-equivariance of $s_H$ and $r_H$). Familiar special
cases of this construction occur when $H$ is a space and $G$ is a
groupoid, so that $H\rtimes G$ is the usual semidirect product
groupoid over $H$ discussed above, and when $G$ and $H$ are both
groups, so that $H\rtimes G$ is the usual semidirect product of
groups.

 Now let $P$ be a  $G$ space. Connes \cite[\S II.10]{Con} notes
 that the tangent bundle $T_G(P)$ of $P$ along $\al$ (i.e.,
 $\ker(\al_*)$, where $\al_*:T(P)\raw T(G^{(0)})$ is the derivative
of $\al$) is a $G$ space, with base map $\xi_p \mapsto\al(p)$
(where $\xi_p\in T_G(P)_p$) and with the obvious push-forward
action. He then regards $T_G(P)$ as a space, and forms the standard
semidirect product groupoid $T_G(P)\rtimes G$ over $T_G(P)$;
to emphasize this, we write the groupoid in question as
\begin{equation}
T_G(P)\rtimes G\stackrel{\raw}{\raw} T_G(P). \label{alain}
\end{equation}
This groupoid is proper, and therefore its \ca\ has computable K-theory.
Connes then defines 
a geometric cycle for $G$ as a proper $G$ space $P$ along with
an element of $$K_*(C^*(T_G(P)\rtimes G\stackrel{\raw}{\raw} T_G(P))).$$

Alternatively \cite{LC}, one could work with the generalized semidirect product
\begin{equation}
 T_G(P)\rtimes G\stackrel{\raw}{\raw} P,\label{klaas}
\end{equation}
where  $T_G(P)$ is seen as a Lie groupoid
over $P$ by inheriting the Lie groupoid structure from $T(P)$
(see Example  \ref{3.1}).
This groupoid fails to be proper, but the following property will be  sufficient.
\begin{Lemma}\label{amenable}
If $P$ is a proper $G$ space, 
then the groupoid  $T_G(P)\rtimes G\stackrel{\raw}{\raw} P$ is amenable.
\end{Lemma}

\textit{Proof.}
Cor.\ 5.2.31 in \cite{AR} states that a (Lie) groupoid $H$ is amenable
iff the associated principal groupoid (that is, the image of the map
$H\raw H^{(0)}\x H^{(0)}$, $h\mapsto (r(h),s(h))$) is amenable and all
stability groups of $H$ are amenable. As to the first condition, the
principal groupoid of $T_G(P)\rtimes G$ is the equivalence relation on
$P$ defined by $p\sim q$ when $q=p\gm$ for some $\gm\in G$.  This is
indeed amenable, because this equivalence relation is at the same time
the principal groupoid of $P\rtimes G\stackrel{\raw}{\raw}P$, which is
proper (hence amenable) because $P$ is a proper $G$ space.  As to the
second condition, the stability group of $p\in P$ in $T_G(P)\rtimes G$
is $T_G(P)_p\rtimes G_p$, where $G_p$ is the stability group of $p\in
P$ in $P\rtimes G$. The former is abelian, and the latter is compact by
the properness of the $G$ action, so that $T_G(P)_p\rtimes G_p$ is
amenable as the semidirect product of two amenable groups. \enp

Despite the fact that the groupoids (\ref{alain}) and (\ref{klaas}) 
are not even  equivalent (in the sense of \cite{MRW}), they have isomorphic \ca s through
a Fourier transformation along the fibers of $T_G(P)$ (seen as a vector
bundle over $P$), and the use of  (\ref{alain}) or (\ref{klaas}) therefore
leads to the same geometric cycles. Hence for the \BCC\ it does not matter
which of these two groupoids is used. However, for the interpretation of the
\BCC\ in terms of deformation quantization one has to work with (\ref{klaas}).
To see this, consider the case where $G$ is trivial. The \ca\ of
(\ref{alain}) is $C_0(T(P))$, which is isomorphic to $C_0(T^*(P))$
through the choice of a Riemannian metric on $P$. On the other hand, 
the \ca\ of (\ref{klaas}) is isomorphic to $C_0(T^*(P))$ through a fiberwise
Fourier transform. It should now be clear from 
Example \ref{3.1} 
that (\ref{klaas}) rather than (\ref{alain}) is the correct groupoid to work with,
if one is interested in relating the \BCC\ to deformation quantization.

Furthermore, 
 the fibered product $P\x_{G^{(0)}} P$ is a $G$ space under the 
base map $(p,q)\mapsto\al(p)=\al(q)$ and the diagonal
action $(p,q)\gm=(p\gm,q\gm)$. Now  $P\x_{G^{(0)}} P$ inherits
a Lie groupoid structure from the pair groupoid $P\x P$ over $P$, 
becoming a Lie groupoid over $P$. Hence one has the semidirect
product groupoid 
$$ (P\x_{G^{(0)}} P)\rtimes G \stackrel{\raw}{\raw}P.$$

The tangent groupoid $G_P$ associated to $P$ has a Lie subgroupoid
$G_P'$ over $I\x P$ that by definition contains all points
$(\hbar=0,\xi_p)$ of $G_P$ whose $\xi_p$ lies in $T_G(P)$, and all
points $(\hbar>0,p,q)$ for which $\al(p)=\al(q)$. It is clear that
$G_P'$ is a field of groupoids over $I$, whose fiber at $\hbar=0$ is
$T_G(P)$, and whose fiber at any $\hbar\in(0,1]$ is $P\x_{G^{(0)}}
P$. Combining the $G$ actions defined in the preceding two cases,
there is an obvious fiberwise $G$ action on $G_P'$ with respect to a
base map $\til{\al}(\hbar,\cdot)=\al_{\hbar}(\cdot)$, where
$\al_{\hbar}=\al_1$ for $\hbar\in(0,1]$. This action is smooth, so
that one obtains a generalized semidirect product groupoid
$$G_P'\rtimes G \stackrel{\raw}{\raw} I\x P.$$ 
This groupoid  is the main tool in the construction of the analytic assembly map occurring
in Connes's version of the \BCC.
\subsection{The analytic assembly map} 
The following lemma provides the continuity conditions tacitly assumed
in \S II.10.$\al$ in \cite{Con}.
\begin{Lemma}\label{conpin1}
If $P$ is a proper $G$ space, then $C^*(G_P'\rtimes G)$ is the 
\ca\ $A$ of sections  of a continuous field of \ca s over $I$ with fibers
\begin{eqnarray}
A_0 & = & C^*(T_G(P)\rtimes G\stackrel{\raw}{\raw} P) ; \nn\\
A_{\hbar} & = & C^*((P\x_{G^{(0)}} P)\rtimes G\stackrel{\raw}{\raw} P)
\:\: \forall\hbar\in(0,1]. \label{CF4}
\end{eqnarray}
This field is trivial away from $\hbar=0$. The same is true if all
groupoid \ca s  are replaced by their reduced counterparts.
\end{Lemma}

\textit{Proof.} 
 The groupoid $G_P'\rtimes G$ inherits the structure of  a smooth field of groupoids over $I$ from the tangent groupoid $G_P$ in the obvious
way. The claim is then immediate from Lemmas \ref{ramlem} and \ref{amenable}.
\enp

 When $G$ is trivial, the continuous field of this proposition is, of course,
the one defined by the tangent groupoid of $P$, which coincides with the field
defined by the Weyl--Moyal deformation quantization of the cotangent bundle $T^*(P)$; 
see Example \ref{3.1}. The general case is a $G$-twisted  version of this,
which cannot really be interpreted in terms of underlying an Poisson manifold,
 because the fiber algebra at $\hbar=0$ is no longer commutative.
\begin{Lemma}\label{conpin2}
The \ca s $C^*((P\x_{G^{(0)}} P)\rtimes G\stackrel{\raw}{\raw} P)$ and
$C^*(G)$ are (strongly) Morita equivalent, as are the corresponding
reduced \ca s.
\end{Lemma}

\textit{Proof.} It is easily checked that the map
$(p,q,\gm)\mapsto\gm$ from $(P\x_{G^{(0)}} P)\rtimes G$ to $G$ is an equivalence
of categories. Since this map is smooth, it follows from Cor.\ 4.23 in
\cite{OBWF} that  $(P\x_{G^{(0)}} P)\rtimes G$ and $G$ are equivalent
as Lie groupoids (and hence as locally compact groupoids with Haar system).
The lemma then follows from Thm.\ 2.8 in \cite{MRW}.
\enp

We have now provided the background for understanding Connes's amazing
construction of the analytic assembly map \cite[\S II.10]{Con}
\begin{equation}
\mu_P: K_*(C^*(T_G(P)\rtimes G))\raw K_*(C_r^*(G)),
\end{equation}
where  $P$ is a proper $G$ space. 
By (\ref{Kmap}),  the continuous field
of Lemma  \ref{conpin1} yields a map  
 \begin{equation} 
 \pi_{1*}\circ\pi_{0*}\inv: K_*(C^*(T_G(P)\rtimes G))\raw K_*(C^*((P\x_{G^{(0)}} P)\rtimes G)).
\label{Cmap}
\end{equation}
 By Lemma \ref{conpin2}
and the fact that the K-theories of Morita equivalent \ca s are isomorphic,
the right-hand side of (\ref{Cmap}) may be replaced by 
 $K_*(C^*(G))$. The canonical
projection $\pi_r$ from $C^*(G)$ to $C_r^*(G)$
pushes forward to $\pi_{r*}:K_*(C^*(G))\raw K_*(C_r^*(G))$,
so that Connes is in a position to define
 \begin{equation}
\mu_P =\pi_{r*}\circ \pi_{1*}\circ\pi_{0*}\inv. \label{Conmu}
\end{equation}

When the classifying space $\underline{E}G$ for proper $G$ actions is a smooth manifold
(which  is true, for example, when $G$ is a connected Lie group
\cite[\S II.10.$\bt$]{Con}, or when $G$ is the tangent groupoid of a manifold),
 the topological K-theory $K^*_{\mathrm{top}}(G)$ is defined as 
 \begin{equation} 
K^*_{\mathrm{top}}(G)= K_*(C^*(T_G(\underline{E}G)\rtimes G)). \label{Ktop}
\end{equation}
 In that case, Connes's analytic assembly map is 
\begin{equation} 
\mu=\mu_{\underline{E}G}:K^*_{\mathrm{top}}(G)\raw K_*(C^*_r(G)).\label{muEG}
 \end{equation}

In general, $K^*_{\mathrm{top}}(G)$ is defined by putting a certain equivalence
relation on the geometric cycles for $G$, and $\mu$ is given by (\ref{Conmu})
applied to each cycle. In any case, the \BCC\ states that $\mu$ should be an isomorphism.
Connes's interpretation 
of this conjecture as a $G$-equivariant version of Bott periodicity
\cite[\S II.10.$\ep$]{Con} is consistent with the 
quantization-oriented approach in this paper, since the field
(\ref{CF4}) underlying the Baum--Connes conjecture is a $G$-twisted
version of the field (\ref{CF1}), which for $M=\R^n$ leads to Bott
periodicity. (See \cite{ENN1,GBV} for a detailed  analysis of the
relationship between Bott periodicity and quantization.)  

Similarly, the usual interpretation of the analytic assembly map as a
generalized index is understandable in the light of the comment below
(\ref{inda}) and a comparison between (\ref{CF1}) and (\ref{CF4}). In
fact, the symbol of a $G$-invariant elliptic pseudodifferential
operator $D$ on $P$ \cite{LMN,NWX} defines an element $[\sg_D]$ of
$K_*(C^*(T_G(P)\rtimes G))$, and the image of this element under
(\ref{Cmap}) is precisely the $K_*(C^*(G))$-valued index of $D$. At
least when $G$ is a group, this argument also bridges the gap between
the usual formulation of the \BCC\ in KK-theory \cite{BCH} and its formulation
due to Connes discussed above, for in that case $D$ defines an
element of the $G$-equivariant K-homology $K_*^G(P)$ of $P$
in terms of which $K^*_{\mathrm{top}}(G)$ is usually defined
(A. Valette, private communication).
\section{Physical interpretation}\label{pi}
\subsection{General comments}
When (\ref{Ktop}) holds, the \BCC\ claims that the $G$-twisted
Weyl--Moyal deformation quantization of the phase space
$T^*(\underline{E}G)$ preserves K-theory. This conjecture is a
far-reaching generalization of the fact that the deformation quantization
of $T^*(\R^n)$ preserves K-theory; as already mentioned, this fact comes down
to Bott periodicity. More generally, Connes's Thom isomorphism
in K-theory \cite{Bla,Con}, which implies Bott periodicity, can be understood through 
deformation quantization
\cite{ENN1}.  The general question whether deformation quantization 
preserves K-theory has been the subject of some research \cite{Nag,Rie5,Ros} 
outside the context of the \BCC, and there are only a few general results. 

We now take a closer look at the continuous field (\ref{CF4}). 
Since the \ca\ $C^*(T_G(P)\rtimes G)$
is noncommutative (unless $G$ is trivial), 
it has no immediate
underlying Poisson manifold, so that  $G$-twisted quantization
cannot itself be seen as quantization.
To analyze the situation, for simplicity we assume that  
$G$ is a Lie group. In that case, the continuous field (\ref{CF4}) 
may be written in terms of conventional crossed
product \ca s \cite{Ped} as 
\begin{eqnarray}
A_0 & = &  C_0(T^*(P))\rtimes G; \nn\\
A_{\hbar} & = &  B_0(L^2(P))\rtimes G
\:\: \forall\hbar\in(0,1]. \label{CF5}
\end{eqnarray}

In the first equation the given $G$ action on $P$ is pulled back first to $T^*(P)$
and subsequently to $C_0(T^*(P))$, and in the second the natural unitary
representation of $G$ on $L^2(P)$ defines an associated action on the
\ca\ $B_0(L^2(P))$ of compact operators by conjugation. We now first  
make  the assumption that  the $G$ action on $P$ is free, allowing a clean analysis,
 to drop it afterwards. 
\subsection{Free actions and superselection theory}
When the $G$ action on $P$ is free (so that $P$ is a principal
$G$ space),  one has a Morita equivalence
\begin{equation}
 C_0(T^*(P))\rtimes G \stackrel{M}{\sim} C_0(T^*(P)/G).\label{ME1}
\end{equation}

This is a special case of a well-known result of Rieffel \cite{Rie00}; in connection
with what follows, another useful proof is to note that 
one has an equivalence of groupoids (in the sense of \cite{MRW})
\begin{equation}
T^*(P)\rtimes G\stackrel{\raw}{\raw} T^*(P)\sim 
 T^*(P)/G \stackrel{\raw}{\raw} T^*(P)/G \label{BB1}
\end{equation}
 through the equivalence bibundle $T^*(P)$.
By \cite{MRW}, this induces  a Morita equivalence of the corresponding
groupoid \ca s, yielding (\ref{ME1}).  

Under the freeness assumption one has an analogous Morita equivalence
on the quantum side, namely 
\begin{equation}
 B_0(L^2(P))\rtimes G
 \stackrel{M}{\sim} C^*((P\x P)/G). \label{ME2}
\end{equation}
 Here $$(P\x P)/G\stackrel{\raw}{\raw}P/G$$ 
is the so-called gauge groupoid of the principal $G$ bundle $P$ \cite{Mac}. 
(When $G$ is compact, the corresponding groupoid \ca\ $C^*((P\x P)/G)$ consists of the $G$-invariant compact operators on $L^2(P)$.) 
To prove (\ref{ME2}), one starts from the equivalence of groupoids 
\begin{equation}
(P\x P)\rtimes G\stackrel{\raw}{\raw}P\x P\sim (P\x P)/G\stackrel{\raw}{\raw} P/G,
\label{BB2}
\end{equation}
 through the equivalence bibundle $P\x P$. Compare (\ref{BB1}).
Thus the Morita equivalent counterpart of the continuous field (\ref{CF5}) is the field
\begin{eqnarray}
A_0' & = &  C_0(T^*(P)/G); \nn\\
A_{\hbar}' & = &  C^*((P\x P)/G)
\:\: \forall\hbar\in(0,1]. \label{CF6}
\end{eqnarray}
This field is continuous as well: in fact, (\ref{CF6})  is just a special case of 
 (\ref{CF3}) in Example \ref{3.4},  in which (with abuse of notation) the groupoid $G$ is taken to be the gauge groupoid $(P\x P)/G$. In particular,  the continuous field (\ref{CF6})
is even a \ca ic deformation quantization of the Poisson manifold $T^*(P)/G$ in the sense of Definition \ref{gsq} (as already mentioned in the Introduction, 
 $T^*(P)/G$ inherits the canonical Poisson structure on $T^*(P)$). 

 Poisson manifolds of this type \cite{Mar}
and their quantization \cite{Lan} have been extensively analyzed.
The underlying classical mechanical system is a particle moving
on the configuration space $Q=P/G$ with an internal degree of freedom
coupling to $G$.  The classical phase space $T^*(P)/G$ decomposes as
a disjoint union of its  symplectic leaves, which may be thought of
as the `classical superselection sectors' of the system. Specifically,
if $J:T^*(P)\raw\mathfrak{g}^*$ is the momentum map of the $G$ action,
the symplectic leaves of $T^*(P)/G$ are the connected components
of the Marsden--Weinstein quotients $J\inv(\mathcal{O})/G$, where
$\CO\subset\mathfrak{g}^*$ is a coadjoint orbit for $G$.
Locally, such a leaf is of the form $T^*(Q)\x\mathcal{O}$. 
The first factor is just the usual phase space of a particle
moving on $Q$, and the second is the classical charge of the particle.
The latter typically couples to an external gauge field \cite{Mar}.

The fact that the quantum algebra of observables
$C^*((P\x P)/G)$ is related to its classical counterpart
$C_0(T^*(P)/G)$  by a \ca ic deformation is  reflected in the superselection
structure of the model. One of Haag's fundamental insights was that
superselection sectors of a quantum system may be identified with
inequivalent irreducible representations of its algebra of observables
 (in quantum field theory further selection criteria are needed, though) \cite{Haa}.
By Lemma \ref{conpin2}, both sides of (\ref{ME2}) 
are Morita equivalent to $C^*(G)$, so that, in particular, 
the superselection sectors of $C^*((P\x P)/G)$ bijectively correspond to the irreducible unitary representations of $G$. Of course, this reflects
 the DHR theory in algebraic quantum field theory \cite{Haa}.
 A comparison with the classical situation then
confirms  Kirillov's  general principle that coadjoint 
orbits should be seen as the classical analogues of 
irreducible unitary representations \cite{Kir}; also cf.\ Example \ref{3.2}. 
\subsection{General actions and singular quantization}
When the $G$ action on $P$ is not free (and this is the main case of
interest in connection with the \BCC), the quotient $T^*(P)/G$ is no longer
a manifold. Nonetheless, its structure is well understood \cite{LMS}.
Each naive symplectic leaf of the form $J\inv(\mathcal{O})/G$
(or rather a connected component thereof)
of $T^*(P)/G$ is not a symplectic manifold, but a 
stratified symplectic space \cite{SL}. In particular, the leaf in
question itself decomposes as a disjoint union of symplectic manifolds,
which are glued together in a certain topological  way that one can describe in detail.
Compared to the regular situation discussed above, this introduces
new classical superselection sectors. 

The problem arises how to quantize such singular symplectic spaces;
cf.\ \cite{Proc} for a survey of what little is known. The
noncommutative geometry approach to the situation would be to
desingularize $T^*(P)/G$ by starting from $C^*(T^*(P)\rtimes G)$
rather than $C_0(T^*(P)/G)$. Although the former \ca\ is
noncommutative, it is still a description of $T^*(P)/G$ as a classical
space. This is reflected by the fact that $A_0=C^*(T^*(P)\rtimes G)$
carries a structure analogous to the notion of a Poisson fibered
algebra defined in \cite{RVW}.  In the \ca ic context, it is necessary
to involve the multiplier algebra to make sense of this idea.

The multiplier algebra of $A_0$ contains $\til{Z}=
C_b^{\infty}(T^*(P))^G$ (where the suffix $G$ denotes the
$G$-invariant functions) in its center, and also contains the
subalgebra $\til{A}_0$ generated by $\til{Z}$ and $\cci(T^*(P)\rtimes G)$.
Then $\til{A}_0$ is a Poisson fibered algebra over $\til{Z}$, in that
one has a bracket $(f,a)\mapsto \{f,a\}$ from $\til{Z}\x\til{A}_0$ to
$\til{A}_0$, which restricts to a Poisson bracket on $\til{Z}$, and is a
derivation on $\til{A}_0$ for fixed $f$ and a derivation on $\til{Z}$
for fixed $a$.  This bracket is simply given by the one on $T^*(P)$,
ignoring the $G$-dependence of $a$.

To quantize the desingularized system, 
one has to deform $C^*(T^*(P)\rtimes G)$. This is precisely what happens in 
Connes's formulation of the \BCC\ described in Section \ref{BCC}.
The continuous field (\ref{CF4}) may be seen as an educated guess to
quantize the singular Poisson manifold $T^*(P)/G$ by 
the \ca\  $B_0(L^2(P))\rtimes G$; the direction of the deformation is
now determined by the more general notion of a Poisson structure
discussed in the previous paragraph. 

This proposal should be tested in concrete examples,
such as the Universe with a Big Bang singularity. 
A complete analysis of this case will have to wait 
for Haag's 90th birthday Festschrift.


\begin{thebibliography}{99}
\itemsep=\smallskipamount
\bibitem{ASZ}  Alvarez, O.,  Singer, I.M.,  B. Zumino, B.: Gravitational
anomalies and the families index theorem.   Commun.\ Math.\
Phys. \textbf{96}, 409--417 (1984).
\bibitem{AG1}   Alvarez-Gaum\'{e}, L.: Supersymmetry and the Atiyah--Singer
theorem.  Commun.\ Math.\ Phys. \textbf{90}, 161--173 (1983). 
\bibitem{AGP}   Alvarez-Gaum\'{e}, L.,   Ginsparg, P.:
The structure of gauge and gravitational anomalies. 
Ann.\ Phys.\ (N.Y.) \textbf{161}, 423--490 (1985).
 \bibitem{AR}
Anantharaman-Delaroche, C.,  Renault,  J.: Amenable groupoids.
 Monographies de L'Enseignement Math\'{e}matique \textbf{36}, 1--196  (2000).
\bibitem{AS1} Atiyah, M.F.,  Singer, I.: The index of elliptic operators I.
Ann.\ Math. \textbf{87}, 485--530 (1968).
\bibitem{At1}  Atiyah, M.F.: Bott periodicity and the index of elliptic operators.
Quart.\ J.\ Math.\ Oxford (2) \textbf{19}, 113--140 (1968).
\bibitem{BC}  Baum, P.,  Connes, A.: Geometric K-theory for Lie groups and
foliations, Ens.\ Math. \textbf{46}, 3--42 (2000). 
\bibitem{BCH}  Baum, P.,   Connes, A.,  Higson, N.: Classifying space for proper actions
and K-theory of group $C^*$-algebras.   Contemp.\ Math.
\textbf{167}, 241--291  (1994). 
\bibitem{Bel}
 Bellissard, J.,  van Elst, A.,  Schulz-Baldes, H.: The noncommutative geometry of the
quantum Hall effect.  J.\ Math.\ Phys. \textbf{35},  5373--5451 (1994).
\bibitem{BGV}  Berline, N.,  Getzler, E.,  Vergne, M.: 
\textit{Heat Kernels and Dirac Operators}. 
 New York: Springer, 1992.
\bibitem{Bis}
   Bismut, J.-M.: Superconnexions, indice local des familles,
d\'{e}terminant de la cohomologie et m\'{e}triques de Quillen.  
M\'{e}m.\ Soc.\ Math.\ France (N.S.) \textbf{46},
27--72 (1991).
 \bibitem{Bla}  Blackadar, B.: \textit{K-theory for Operator Algebras}, 2nd ed. 
 Cambridge: Cambridge University Press,  1999.
\bibitem{Blan}   Blanchard, \'{E}.: D\'{e}formations de $C^*$-alg\`{e}bres 
de Hopf, Bull.\ Soc.\ math.\ France \textbf{124}, 141--215  (1996).
\bibitem{Cad}  Cadet, F. : \textit{D\'{e}formation et Quantification par Groupo\"{\i}de des
Vari\'{e}t\'{e}s Toriques}.  Ph.D.\ thesis,
Universit\'{e} d'Orl\'{e}ans, 2001.
\bibitem{CW}
Cannas da Silva,  A., Weinstein, A.: \textit{Geometric Models for Noncommutative 
Algebras}.  Providence (RI): Amer.\ Math.\ Soc.,   1999.
\bibitem{CCFGRV}   Cari\~{n}ena,  J.F.,   Clemente-Gallardo, J.,
  Follana, E.,   Gracia-Bond\'{\i}a, J.M.,   Rivero,  A.,    V\'{a}rilly, J.C.:
Connes' tangent groupoid and strict quantization.
 J.\ Geom.\ Phys. \textbf{32},  79--96 (1999). 
\bibitem{CEN}
   Chabert, J., Echterhoff, S.,  Nest, R.:
The Connes--Kasparov conjecture for almost connected groups. 
\texttt{arXiv:math.OA/0110130}.
\bibitem{CF} Crainic, M., Fernandes, R.L.: 
                Integrability of Lie brackets. Ann.\ Math.,
                to appear. \texttt{arXiv:math.DG/0105033}.
\bibitem{Con}   Connes, A.: \textit{Noncommutative
                Geometry}. San Diego: Academic Press,  1994.
\bibitem{CH}   Connes, A.,   Higson, N.: D\'{e}formations, morphismes asymptotiques
et K-th\'{e}orie bivariante. C.R.\ Acad.\ Sci.\ Paris S\'{e}r.\ I Math.
\textbf{311}, 101--106 (1990).
\bibitem{CL}   Connes A.,    Landi, G.:
Noncommutative manifolds, the instanton algebra and isospectral deformations.
Commun.\ Math.\ Phys. \textbf{221}, 141--159  (2001).
\bibitem{CDW}
 Coste, A.,  Dazord, P.,  Weinstein,  A.: Groupo{\"\i}des symplectiques.
 Publ.\ Dept.\ Math.\ Univ.\ Lyon 1 \textbf{2A},    1--62 (1987).
\bibitem{Cou}   Courant,  T.J.: Dirac
Manifolds.  Trans.\ Amer.\ Math.\ Soc. \textbf{ 319} (1990) 631--661.
\bibitem{Dix}   Dixmier,  J.: \textit{$C^*$-Algebras}. Amsterdam: North--Holland,   1977.
\bibitem{ENN1}    Elliott, G.A.,   Natsume, T.,   Nest,  R.:
The Heisenberg group
and K-theory.  K-theory \textbf{7}, 409--428 (1993).
\bibitem{ENN2}   Elliott, G.A.,   Natsume, T.,   Nest,  R.:
The Atiyah--Singer index theorem as passage to the classical limit in
quantum mechanics.  Commun.\ Math.\ Phys. \textbf{182},
505--533 (1996).
 \bibitem{Fed}  Fedosov,  B.: \textit{Deformation Quantization and Index Theory}.
Berlin: Akademie-Verlag, 1996.
\bibitem{FW}  Friedan  D.,   Windey,  P.:
Supersymmetric derivation of the
Atiyah--Singer index and the chiral anomaly. Nucl.\ Phys. \textbf{B235},
395--416 (1984).
\bibitem{GBV}   Gracia-Bond\'{\i}a,  J.M. ,   V\'{a}rilly, J.C.,   Figueroa,  H.:
\textit{Elements of Noncommutative Geometry}. Boston: Birkh\"{a}user,     2001.
 \bibitem{Haa} Haag, R.: \textit{Local
Quantum Physics}.  2nd ed. Berlin: Springer, 1996.
 \bibitem{Hig}  Higson,  N.: On the K-theory proof of the index theorem.
Contemp.\ Math. \textbf{148}, 67--86 (1993). 
 \bibitem{HS}
 Hilsum M.,      Skandalis,  G.:  Morphismes $K$-orient\'{e}s d'espaces de feuilles et
fonctorialit\'{e} en th\'{e}orie de Kasparov (d'apr\`{e}s une conjecture d'A. Connes).
Ann.\ Sci.\ d'\'{E}cole 
Norm.\ Sup.\ (4)  \textbf{20},  325--390  (1987).
\bibitem{KP}
Kellendonk, J.,  Putnam, I.F.: Tilings, $C^*$-algebras, and
$K$-theory. In:  \textit{Directions in Mathematical Quasicrystals}. CRM
Monogr. Ser., \textbf{13}. Providence (RI): Amer. Math. Soc., 2000, pp.\ 177--206.
\bibitem{KW}   Kirchberg,  E.,  Wassermann,   S.:
Operations on continuous bundles of 
$C^*$-algebras.  Math.\ Ann. \textbf{303}, 677--697  (1995).
\bibitem{Kir} Kirillov, A.A.: \textit{Elements of the Theory of
Representations}. Berlin: Springer, 1976.
\bibitem{Lan} Landsman,  N.P.: \textit{Mathematical Topics Between
Classical and Quantum Mechanics}.     New York: Springer,  1998.
\bibitem{LGCA}  Landsman,  N.P.:  Lie groupoid $C^*$-algebras and Weyl quantization.
 Commun.\  Math.\ Phys. \textbf{206}, 367--381  (1999). 
 \texttt{arXiv:math-ph/9807028}.
\bibitem{NPLOA} Landsman,  N.P.:   Operator algebras and
Poisson manifolds associated to  groupoids.
 Commun.\  Math.\ Phys.
\textbf{222},  97--116 (2001). 
\texttt{arXiv:math-ph/0008036}.
\bibitem{OBWF}  Landsman,  N.P.: Quantized reduction as a tensor product.
In: Ref.\ \cite{Proc}, pp.\ 137--180.
\texttt{arXiv:math-ph/0008004}. 
\bibitem{LR} Landsman,  N.P.,      Ramazan,  B.: Quantization of Poisson 
algebras associated to  Lie algebroids.  Contemp.\
 Math. \textbf{282}, 159--192 (2001). 
 \texttt{arXiv:math-ph/0001005}.
\bibitem{LC} Landsman,  N.P.: Quantization and the tangent groupoid. In:
Cuntz, J.\ (ed.), \textit{Proc.\ 4th Operator Algebras Int.\ Conf.: Operator Algebras and 
Mathematical Physics (Constantza, 2001)}, to appear.  \texttt{arXiv:math-ph/0208004}. 
\bibitem{Proc} 
 Landsman, N.P., Pflaum,   M.,  Schlichenmaier,  M.  (eds.):
\textit{Quantization of Singular Symplectic Quotients}. 
Basel: Birkh\"{a}user,  2001.   
\bibitem{LMN} Lauter, R.,  Monthubert, B.,  Nistor, V.: Pseudodifferential analysis on
continuous family groupoids. Doc.\ Math. \textbf{5}, 625--655 (2000).
\bibitem{LMS} Lerman, E.,  Montgomery, R., Sjamaar, R.: 
Examples of singular reduction.  In: Salomon, D. (ed.), \textit{Symplectic
Geometry}. LMS Lecture Notes Series \textbf{192}.
 Cambridge: Cambridge University Press, 1993, pp.\ 127--155.
 \bibitem{Lin}   Lin, H.: \textit{An Introduction to the Classification of Amenable C*-Algebras}.
  Singapore: World Scientific, 2001.
\bibitem{Mac}  Mackenzie,   K.: \textit{ Lie Groupoids and Lie Algebroids in
Differential Geometry}. Cambridge: Cambridge University Press,  1987.
second edition, ibid., to appear.
\bibitem{Mar} Marsden, J.E.: \textit{Lectures on Mechanics}. Cambridge:
 Cambridge University Press, 1992.
\bibitem{MRW}   Muhly, P.,  Renault, J.,    Williams, D.: Equivalence and 
isomorphism for groupoid \ca s.  J.\ Operator Th.\ \textbf{17}, 3--22  (1987).
\bibitem{Nag}   Nagy, G.: Deformation quantization and K-theory. 
Contemp.\ Math. \textbf{214}, 111--134 (1998).
\bibitem{NT1}  Nest,  R.,   Tsygan, B.: Algebraic index theorem.  Commun.\ Math.\ Phys.
\textbf{172}, 223--262 (1995).
\bibitem{NWX}
 Nistor, V.,  Weinstein, A.,  Xu, P.: Pseudodifferential operators on 
differential groupoids.  Pac.\ J.\  Math.  \textbf{189},  117--152 (1999).
\bibitem{Ped} Pedersen, G.K.: \textit{
$C^*$-Algebras and their Automorphism Groups}. London: Academic Press, 1979.
\bibitem{Pflaum}  Pflaum,  M.J.: A deformation-theoretical approach to Weyl quantization on
Riemannian manifolds.  Lett.\ Math.\ Phys. \textbf{45},  277--294  (1998).
 \bibitem{Ram}    Ramazan,  B.: \textit{
Deformation Quantization of Lie--Poisson Manifolds}.  Ph.D.\ thesis,
Universit\'{e} d'Orl\'{e}ans, 1998.
\bibitem{Ren}
  Renault, J.: \textit{A Groupoid Approach to C*-algebras}.  Lecture Notes in
Mathematics \textbf{793} (1980).
\bibitem{RVW} Reshetikhin, N., Voronov, A., Weinstein, A.:
Semiquantum geometry. 
J.\ Math.\ Sci. \textbf{82}, 3255--3267 (1996).
\bibitem{Rie00}
Rieffel, M.A.:  Applications of strong Morita equivalence to
transformation group $C\sp{*} $-algebras. In: Kadison, R. (ed.), 
\textit{Operator Algebras and
Applications, Part I}. Proc.\ Sympos.\  Pure Math. \textbf{38}. 
Amer. Math. Soc., Providence (RI), 1982,
 pp. 299--310.
\bibitem{Rie0} Rieffel, M.A.:
 Continuous fields of $C^*$-algebras coming from group 
cocycles and actions.  Math.  Ann. \textbf{283},  631--643 (1989).
\bibitem{Rie1}  Rieffel, M.A.:
Deformation quantization of Heisenberg manifolds.  Commun.\ Math.\
Phys. {\bf 122}, 531--562 (1989).
\bibitem{Rie2}  Rieffel, M.A.: Quantization and
$C^*$-algebras.  Contemp.\ Math. \textbf{167}, 
67--97  (1994). 
\bibitem{Rie4} Rieffel, M.A.:
  Lie group convolution algebras as deformation quantization 
of linear Poisson structures.  Amer.\ J.\ Math. \textbf{112},    657--686 (1990).
\bibitem{Rie5} Rieffel, M.A.:
  $K$-groups of $C\sp *$-algebras deformed by actions of $R^d$. J.\
Funct.\ Anal. \textbf{116},   199--214 (1993) .
\bibitem{Ror}   R\o rdam, M.,  Larsen,  F.,
                   Laustsen, N.:  \textit{An introduction to K-theory for C*-algebras}. Cambridge:
Cambridge University Press, 2002. 
\bibitem{RS}  R\o rdam, M., St\o rmer, E.:  \textit{
Classification of Nuclear $C^*$-algebras. Entropy in Operator
Algebras}.  Encyclopaedia of Mathematical Sciences:
                  Operator Algebras and Non-commutative Geometry.  Heidelberg:  Springer, 2002. 
\bibitem{Ros}   Rosenberg, J.: Behaviour of K-theory under quantization. In:
Doplicher, S. et al.\ (eds.),
\textit{Operator Algebras and Quantum Field Theory}. Boston:  International Press,  
1996, pp.\ 404--413. 
\bibitem{SL}
Sjamaar R.,  Lerman, E.: Stratified symplectic
spaces and reduction. Ann.\ Math.  \textbf{134}, 375--422 (1991).
\bibitem{Ska} Skandalis, G.: Progr\`{e}s r\'{e}cents sur la conjecture de Baum--Connes.
 Contribution de Vincent Lafforgue.
Ast\'{e}risque  \textbf{276},   105--135  (2002). 
\bibitem{Tu}   Tu, J.-L.: The Baum--Connes conjecture for groupoids. In:
 \textit{$C^*$-algebras (M\"{u}nster, 1999)}. Berlin, Springer,  2000, pp.\ 227--242. 
\bibitem{Val}  Valette,  A.: \textit{Introduction to the Baum--Connes conjecture}.
 Basel: Birkh\"{a}user, 2002.
\bibitem{Vor}   Voronov, T.: Quantization on supermanifolds and the analytic proof of the
Atiyah--Singer index theorem.  J.\ Soviet Math. \textbf{64}, 993-1069 (1993).
\bibitem{Wei89}  Weinstein, A.: Blowing up realizations of
Heisenberg-Poisson manifolds.    Bull.\ Sc.\ math.  \textbf{(2) 113}, 
381-406 (1989).
\bibitem{WitK} Witten, E.: D-branes and K-theory. \textit{JHEP} \textbf{98}12:019 (1998).
\end{thebibliography}
\end{document}